\documentclass[aps,prl,a4paper,10pt,showpacs,twocolumn]{revtex4}

\usepackage{amsmath}
\usepackage{amssymb}
\usepackage{graphicx}
\usepackage{xcolor}
\usepackage{braket}
\usepackage{hyperref}
\usepackage{wrapfig}

\newcommand{\vb}[1]{{\boldsymbol {#1}}}
\newcommand{\ud}{\mathop{}\!\mathrm{d}} 

\usepackage{dcolumn}

\begin{document}

\title{
        Coherent Amplification of
                Ultrafast
                Molecular
                Dynamics
                in an Optical Oscillator
      }
\author{Igal Aharonovich}
\email{jigal2@gmail.com}
\author{Avi Pe'er}
\email{avi.peer@biu.ac.il}
\affiliation{Department of Physics and BINA center for nano-technology, Bar-Ilan University, Ramat Gan 52900, Israel}

\date{\today}

\begin{abstract}
Optical oscillators present a powerful optimization mechanism. The inherent competition for the gain resources between possible
modes of oscillation entails the prevalence of the most efficient single mode.
We harness this 'ultrafast' coherent feedback to optimize an optical field \emph{in time}, 
and show that when an optical oscillator based on a molecular gain medium is synchronously-pumped by ultrashort pulses,
a temporally coherent multi mode field can develop that optimally dumps a general, dynamically-evolving vibrational
wave-packet, into a \emph{single vibrational target state}.
Measuring the emitted field opens a new window to visualization and control of fast molecular dynamics.
The realization of such a coherent oscillator with hot alkali dimers appears within experimental reach.
\end{abstract}

\pacs{42.50.-p, 42.50.Md, 42.55.Ye}
\keywords{Coherent Raman Laser; Molecular vibrational dynamics}

\maketitle

Compared to atoms, molecules are unique in their vibration and rotation, which produces rich, complex dynamics upon excitation by pulses of light \cite{herzberg_1989}.
Since chemical reactions are driven by vibrational dynamics \cite{atkins2011molecular},
precise measurement and control of the vibration is of interest.
To date, vibrational dynamics was measured using either pump-probe excitation or wave-packet tomography.
In pump-probe, a pump pulse excites the dynamics and a delayed probe pulse probes it,
usually by selective ionization or dissociation of the molecule depending on its vibrational configuration \cite{Katsuki17032006, pmid19392112, PhysRevA.43.5153}.
In wave-packet tomography, the spontaneous emission from an excited vibrational wave-packet is time-resolved,
reflecting the vibration dynamics due to the time-dependent Franck-Condon overlap between the
wave-packet and the final state \cite{0953-4075-31-9-004, PhysRevLett.80.1418, PhysRevLett.96.103002, RIS_0}.
Both methods are inherently limited by the weak measured signal (less than one photon or electron per molecule per pump pulse),
which is difficult to detect, indicating that both methods are inherently slow and require averaging
(either in space over a large ensemble of molecules or in time over many pump pulses) to obtain a decent signal-to-noise.
While averaging incoherently accumulates the light intensity from many molecules,
we present a method for \emph{coherent} accumulation of the field \emph{amplitude} in time
that can dramatically improve the signal-to-noise and speed of the measurement.

\begin{figure}[tbp]
    \centering
    \includegraphics[width=\linewidth]{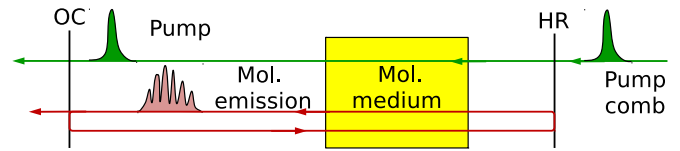}
    \caption{
             Oscillator concept. A molecular medium in an optical cavity is excited by a train of pump pulses. Each pump pulse launches a vibrational wave-packet, 
             which later emits (Raman) light, as it vibrates on the excited molecular potential.
             The cavity, which resonates only the Raman emission (not the pump), is
             matched to the repetition rate of the pump pulses, allowing amplification of the Raman emission.
             }
  \label{fig:apparatus}
\end{figure}

\begin{figure}[tbp]
    \centering
    \includegraphics[width=\linewidth]{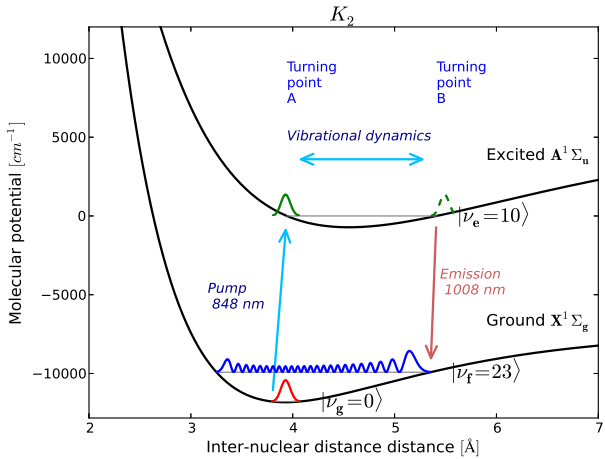}
    \caption{
             Molecular excitation cycle for the $K_2$ dimer.
             The pump pulse excites a vibrational wave-packet.
             As the wave-packet vibrates and disperses on the excited potential,
             emission occurs primarily when the wave packet passes at the outer turning point B,
             where the Franck-Condon overlap is maximal to the target state (near vibrational level $\nu_{f} = 23$).
             }
    \label{fig:potential_curves}
\end{figure}

We suggest an optical oscillator, where coherent emission from a dynamically evolving wave-packet
(excited by an ultrashort pump pulse) is amplified beyond the oscillation threshold. The oscillator concept is outlined in figure \ref{fig:apparatus}:
A medium of molecules is placed in an optical cavity and excited by ultrashort pump pulses
with a repetition rate that matches the cavity round trip. Each pump pulse excites a non-stationary
vibrational wave-packet, which later coherently evolves (vibrates) on the excited electronic potential.
During vibration, the wave-packet emits non-stationary Raman light, either spontaneous or stimulated,
as illustrated in figure \ref{fig:potential_curves} for a medium of $K_2$ molecules. In wave-packet tomography
this spontaneous Raman emission was temporally resolved to reconstruct the wave-packet dynamics
\cite{Katsuki17032006,walsmsley_1996_book,0953-4075-31-9-004}. We suggest to collect the emitted Raman light in a cavity,
where stimulated amplification can occur when the pump repetition matches the cavity round trip,
which ensures that the emission from one pump pulse returns to the medium synchronously with the next pulse. When such a Raman amplifier crosses threshold,
a temporally coherent oscillation can be obtained, where the pump \emph{field} that was inscribed onto the excited wave-packet,
is later reshaped by the vibrational dynamics and re-emitted in the form of coherent Raman radiation.
For simplicity we assume that the pulse repetition is low compared to the decay rate of the molecules,
indicating that the time between pulses is sufficiently long for all molecules to decay back to the ground vibrational level.
The molecules therefore carry no memory of previous excitations, but the light-field accumulated in the cavity serves as coherent memory that lingers from pulse to pulse.

Our method is related to work on precision control of molecular dynamics using the frequency comb
\cite{PhysRevLett.98.113004,PhysRevLett.101.023601,Stowe20081},
where highly efficient and selective population transfer was achieved between designated vibrational levels,
relying on coherent accumulation of molecular excitations from a train of weak pump-dump pulse-pairs,
and a vibrational wave packet as an intermediate.
The noticeable difference is that now the coherent memory is not in the molecules, but in the accumulated dump field,
and the field is not pre-specified, but rather amplified from spontaneous emission. Yet, the logic of coherent accumulation is identical,
leading to selective transfer, and the theoretical treatment is similar, both for analytic calculation and numerical simulation of the dynamics.

Note the difference between the proposed coherent Raman oscillator and methods of coherent Raman spectroscopy \cite{JRS:JRS4221,coherent_raman_micrscopy_2012},
such as Raman fluorescence, stimulated Raman \cite{Freudiger2011,Saar03122010}
and coherent anti-Stokes Raman spectroscopy \cite{Pezacki2011,:/content/aip/journal/apl/80/9/10.1063/1.1456262,Dudovich2002}.
All these methods measure molecular vibrational levels in the \emph{ground electronic potential},
and deliberately \emph{avoid the excited potential} to ensure a purely virtual Raman transition between ground levels.
In contrast, we aim to observe the vibrational dynamics in the \emph{excited electronic potential}, and the pump pulses
are tuned to excite a meaningful wave-packet.
In addition, the dump pulse is not externally set, but rather amplified from spontaneous emission through coherent accumulation and is subject to mode-competition.

In the analytic and numerical study presented here we consider a realistic test case of a coherent Raman oscillator.
As gain-medium we take an ensemble of alkali dimers ($K_2$ or $Li_2$) that are thermally mixed with free atoms
in a hot vapor cell ($\sim\!550^{\circ} K$ for $K_2$, and $\sim\!700^{\circ}K$ for $Li_2$),
and show that the oscillation threshold is achievable with reasonable pump power ($\sim\!1$W),
molecular densities ($10^{13-14}\text{cm}^{-3}$ for $K_2$) and interaction length ($5\!-\!10\text{cm}$).
We present the calculation and simulation in a bottom-up structure, from the microscopic,
single molecule dynamics to the macroscopic field gain and cavity evolution, accounting for loss, dispersion and decoherence.
Since the performance of this coherent oscillator, and in particular the oscillation threshold,
critically depend on decoherence properties, an in-depth evaluation of decoherence mechanisms is
provided - both homogeneous pressure broadening (collisions) and inhomogeneous rotational and Doppler broadenings.
We show that the homogeneous coherence time in a hot vapor cell can reasonably be $T_2\!\geq\!100$ps,
and that the threshold is primarily affected by the inhomogeneous thermal distribution of rotational states,
which breaks the molecular ensemble into independent coherent clusters and reduces the available population for
coherent gain. This letter outlines the calculation concept, simulation procedure and results,
and the decoherence considerations (with full details provided in the additional online material \ref{sec:supplementary_material_crl}).

Before dwelling into the calculation, it is illuminating to present the unique features of this oscillator as
reflected in the simulation results. Once Raman oscillation is obtained, this oscillator demonstrates unique coherent dynamics:
The produced 'dump' field, that stimulates the molecules back to the ground potential,
forms together with the pump pulse a coherent pump-dump pair, reminiscent of many configurations of coherent control
\cite{shapiro_2011,:/content/aip/journal/jcp/83/10/10.1063/1.449767}.
Here however, the dump field is not specified a-priori, but is dynamically amplified from spontaneous emission.
The final target state of the molecules is therefore also undefined, and different possible decay channels may compete for the gain (pump energy). 
Surprisingly, the winning decay channel near threshold is to dump the entire wave-packet to a \emph{single}
target vibrational state with a train of dump pulses that is matched to the vibrational period of the
excited wave-packet (see figure \ref{fig:simulation_results_near_threshold}). Furthermore, the quantum efficiency of the dump transfer is always
\emph{near unity} with very high target selectivity, even very close to threshold.

The oscillator harnesses mode-competition to 'automatically' solve an important optimization problem of coherent quantum control: to find the optimal
dump pulse for efficient and selective transfer from a given wave-packet to a single target state
\cite{PhysRevLett.98.113004,PhysRevLett.101.023601,:/content/aip/journal/jcp/84/7/10.1063/1.450074}.
This automatic solution for the pulse in time is in direct analogy to work in \cite{NaturePhoton.7.919},
where the competition between spatial modes was exploited to generate the optimal field \emph{in space} to
focus through a highly-scattering, turbulent medium.
Single vibrational mode selection was also observed in a theoretical study of a proposed X-ray
laser with a gain medium of $N_2$ molecules \cite{1742-6596-488-3-032019,PhysRevLett.110.043901}.

Although the final state is not designated a-priori, and is selected via intra-cavity mode competition, the dynamics can be steered
towards a desired vibrational target state $n_f$ by shaping the spectrum of the pump pulse to maximize the spectral overlap of
the excited wave-packet with the target state \cite{PhysRevLett.98.113004}.
As detailed in the online supplementary material \ref{sec:supplementary_material_crl}, this is accomplished by enhancing (diminishing) pump frequencies that excite components of the wave-packet with
high (low) Franck-Condon overlap to the target state $n_f$. Specifically, shaping the pump spectrum according to $E_{\text{pump}}(\omega_{n_e}) \propto \mu_{eg}  \braket{n_e|n_f}$,
where $\omega_{n_e}$ denotes the transition frequency from the initial ground state $n_g\!=\!0$ to the vibrational component $n_e$ of the excited wave-packet.
We verified this selective steering in simulation with $Li_2$, where shaping the spectrum of a given pump-pulse steered the dump into any of the target states $n_f\!=\! 8, 9, 10$.

One can consider the emission from the coherent Raman oscillator as a Raman shifted version of optical free induction decay (FID),
where a large ensemble of molecular dipoles emit coherently.
Observation of FID in the optical domain is challenging since the emission is very short in time (limited by the coherence time $T_2$),
and overlaps spectrally with the excitation pulse.
Previously, optical FID was measured indirectly by Fourier inversion of the molecular absorption spectral amplitude (including phase),
as measured with a dual frequency comb \cite{PhysRevLett.100.013902,Coddington:10}. The large Raman shift in our oscillator allows optical FID to be observed directly in time.
Furthermore, the excited wave-packet dynamics can be directly read-off from the spectrum and phase of the emitted field,
as elaborated in the supplementary material.

To model the coherent Raman oscillator we take an iterative approach, which mimics the intra-cavity evolution in time: the molecular
radiated field in every round trip is calculated based on the the so-far accumulated intra-cavity field and on the recurring pump pulse excitation,
both interacting coherently with the molecular ensemble. The cavity field for the next round trip is then
generated by adding the emission to the previous intra-cavity field, including decoherence, cavity-loss and dispersion.
The emitted field $E_{em}$ is proportional within the dipole approximation to the $2^{\text{nd}}$
time-derivative of the total (macroscopic) electric dipole in the medium $P_M$
\begin{equation} \label{eq:emitted_field}
    E_{em}(z,t)\!=\!\dfrac{1}{4\pi z \varepsilon_0 c^2}\dfrac{\ud^2}{\ud t^2}P_M(t\!-\!z/c).
\end{equation}
In a large ensemble, the total dipole is $P_M(t)\!=\!\mathcal{N}P(t)$, where $\mathcal{N}$ is the effective number of molecules
and $P$ is the quantum average dipole of a single molecule:
\begin{equation} \label{eq:average_dipole}
        P(t)\!=\! e^{- i \omega t}\mu_{eg}\braket{ \psi_{g}(t) | \psi_{e}(t) }\!+\!cc\! \equiv\! p(t) e^{-i \omega t}\!+\! cc.
\end{equation}
Here $\ket{\psi_{g,e}(t)}$ are the vibrational wave-functions on the
ground and excited electronic potentials, $\omega$ is the center optical frequency and $\mu_{eg} = \bar{\mu}_{ge} \equiv \mu$
is the electronic dipole moment between the potentials, assumed independent of the inter-nuclear distance (Condon approximation).
Thus, once the molecular wave-packet dynamics is calculated, the microscopic molecular dipole (and emitted field)
can be easily obtained as the time dependent overlap between the excited and target wave-packets (and its temporal derivatives).

The gain is thus calculated in three stages: First, given the pump pulse field, and the current cavity-accumulated dump field,
the dynamics of the wave packet is calculated on all the coupled ground-excited-target potentials by solving numerically
the time-dependent Schr\"{o}dinger equation (using the split-operator method \cite{0034-4885-58-4-001}).
Once the wave-functions are calculated in time, the average dipole and the emitted field (per molecule)
are calculated with equations \ref{eq:emitted_field} and \ref{eq:average_dipole}.
Last, the macroscopic field gain is calculated by considering the spatial mode of the emitted field and the total number of excited molecules.
Cavity-losses and dispersion are applied after every iteration, and spontaneous emission is modeled as an
additive white noise to the emitted field, which seeds the oscillation.
Homogeneous decoherence (collisional pressure broadening) is introduced as a phenomenological decay of the macroscopic dipole,
whereas inhomogeneous decoherence is incorporated as an effective reduction of the available molecular population due to the
thermal spread of rotations ($J$ states), which divides the molecular population to independent coherent clusters of slightly
shifted emission frequencies (see further details later on and a complete discussion in the supplementary material).

The evolution of the Raman oscillator can then be simulated including all experimental parameters,
the mode competition during cavity buildup can be fully visualized for both the wave-packet dynamics
and the cavity field, and the emitted field in stable operation can be calculated.
We simulated the coherent Raman oscillator for molecular media of alkali dimers $Li_2, K_2$ and $Rb_2$,
using electronic Morse potential fits \cite{PhysRevA.54.204,Jasik2006563,Park2001129}
for the ground $X^{1} \Sigma_{g}$ and excited $A^{1} \Sigma_{u}$ states.
Transition dipole moments were assumed to be the spherically averaged atomic values of the D-line
\cite{gehm_properties_of_lithium}.

\begin{figure}[tbph]  
    \centering
      \includegraphics[width=0.95\linewidth]{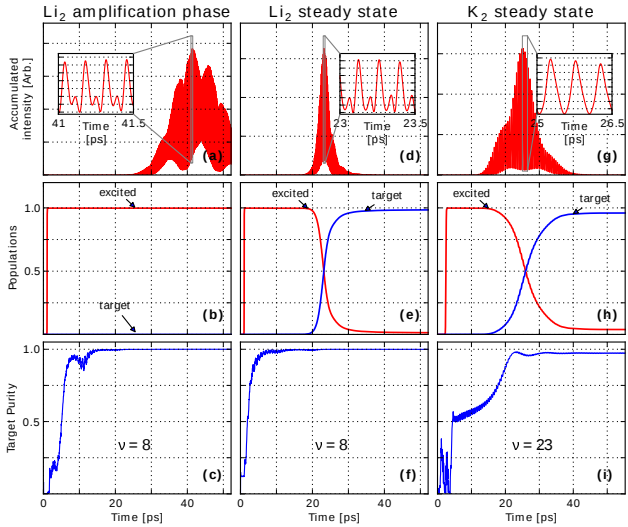}
      \caption
      {
                Simulation results (for a single coherent cluster)
                for $Li_2$ and $K_2$ during 50ps following a particular pump pulse excitation  ,
                for an oscillator pumped \emph{slightly above} threshold.
                The left column shows the intra-cavity field and populations at an early stage of the amplification, well before saturation
                is reached ($Li_2$);
                the middle ($Li_2$) and right ($K_2$) columns show the results for stable oscillation.
                The top row shows the temporal intensity of the intra-cavity field,
                where the rapid oscillations (see inset) match with the vibrational dynamics of the excited wave packet.
                The middle row shows the corresponding evolution of the populations of the excited (red) and target (blue) wave packets.
                In the bottom row, the overlap of the target state with the designated target state is shown, demonstrating the high selectivity of the dump transfer.
      }
    \label{fig:simulation_results_near_threshold}
\end{figure}

Figure \ref{fig:simulation_results_near_threshold} shows snapshots of the accumulated field, the molecular populations
and the vibrational selectivity for $Li_2$ and $K_2$ dimers for various stages of oscillator evolution (with a single oscillating coherent cluster).
The results reveal exceptional features of this coherent Raman oscillator:
(a) \emph{Complete dumping} is obtained of the excited wave-packet to the target state,
even though the oscillator is pumped just a few percent above threshold. This is in contrast to standard lasers,
where the excited state population clamps to the threshold value, and can never be completely dumped.
(b) Near threshold, the emitted field forms a long train of pulses matched to the excited wave-packet vibration
(see insets in the top row of figure \ref{fig:simulation_results_near_threshold}).
Note that the main emission develops only $\sim\!30ps$ after the pump in spite of the shorter $T_2\!=\!25$ps that was assumed.
%
(c) \emph{The vibrational selectivity of the target state is exceptional} near threshold. For $Li_2$,
practically all the dumped population occupies a single vibrational state ($> 99.99\%$),
and this selectivity is achieved rather early in time, even before the main bulk of the population is actually transferred.
For $K_2$ selectivity is a little lower at $> 98\%$, and for $Rb_2$ the Selectivity was $> 90\%$.
These high selectivity values are achieved autonomously by the system due to the physical preference in the mode competition stage.
This preference is directly related to the ratio of the excited vibrational period to the available coherence time $T_2$, which is best for the light and fast $Li_2$.
(d) As the pump is increased further above threshold, the emitted pulse train becomes shorter and appears
at an earlier time after the pump. The selectivity of the target state in $K_2$ is also gradually degraded,
showing mixture of states (see supplementary material \ref{sec:supplementary_material_crl}).
$Li_2$ shows a similar trend, but its selectivity is much more robust and is degraded only at much higher, rather impractical pump levels.
(e) Conversely, as the pump is reduced towards threshold, the main oscillation is pushed towards longer times,
until eventually suppressed by the decoherence window and can no longer dump all the excited population to the target state.
\emph{The threshold oscillation is therefore a direct result of the available coherence time, and if longer coherence is assumed,
the threshold would be reduced}, as indeed observed in our simulations.

To explain the above properties, we consider that just like any laser, the coherent Raman oscillator
'seeks' the most efficient channel to dump the excited population.
Due to the coherent pumping, the oscillator can exploit coherent population transfer which the standard laser cannot (e.g., a $\pi$-pulse).
Specifically, by extending the coherent transfer over a longer time, a complete dump transfer can be achieved with a lower overall dump energy,
just like in a simple two-level system, where the energy required for a $\pi$-pulse is inversely proportional to its duration.
Thus, near threshold, where the available energy for the dump is low, a long coherent train (limited by decoherence) develops.
As the pump is increased above threshold, the available dump energy increases,
and the population can be dumped faster.

The molecular dynamics however is far richer than that of a two-level system, and both the excited and the target states are generally time dependent
wave-packets that vibrate on two different potentials. Thus, in order for a coherent transfer to occur over many vibrational periods, some form of a dynamical
relation must be met: One option is that the two wave-packets will vibrate 'in unison', allowing a dynamic coherent transfer 'as they move'.
Since the two wave-packets have different vibrational periods, the duration of such a dynamic transfer is inherently limited by the
vibrational frequency difference between the two potentials $\tau_{\text{dyn}}\! <\! 1 / (\nu_{e}\! -\! \nu_{t})$.
Prolonging the coherent transfer beyond $\tau_{\text{dyn}}$, is only possible if the dynamics of the target state can be 'frozen',
such that it remains stationary over the entire transfer, i.e., have it an eigen-state of the target potential.
Thus, near threshold, where the transfer is slow, the oscillator tends to select a single target state,
whereas far above threshold, where the transfer duration is shortened, dynamical transfer is allowed and the selectivity of the target state is reduced.

Eventually, the coherent transfer duration is limited by the available coherence time $T_2$.
Consequently,
to achieve a single target state, it is necessary that $T_2$ be long enough to allow target state selectivity, i.e.,
\begin{math} \label{eq:selectivity_criterion}
    T_2 > \frac{1}{\nu_{e} - \nu_{t}}.
\end{math}
This can explain the differences between $Li_2$, $K_2$ and $Rb_2$ in target selectivity: The coherence time in our simulation was fixed
at $T_2 = 25 ps$. 
For the light $Li_2$, the vibrational frequencies (and their differences) are high,
so the selectivity criterion
is well met.
$K_2$ is heavier and therefore the requirement
is only marginally fulfilled,
yet good selectivity can be obtained near threshold. $Rb_2$ on the other hand, is too heavy and falls short of this criterion.
If the coherence time in the experiment will be longer, $K_2$ and $Rb_2$ could also show high selectivity, in addition to a reduced threshold.

To calculate the threshold, it is now necessary to estimate the homogeneous coherence time $T_2$ in the vapor cell,
dictated by the collisional pressure broadening, and the available density for coherent gain, affected by the
inhomogeneous distribution of rotational states. Since the pressure broadening for collisions of the molecules
with the surrounding atoms (and buffer gas, if added) is of order 100MHz/Torr, $T_2\!\geq\!100$ps is easily
achievable at atomic pressures up to 10Torr.
For inhomogeneous broadening, the major limitation is due to rotations, which are thermally populated up to $J\!\approx\!100$.
The optical emission frequency of each $J$ state is slightly shifted due to the difference in rotational constants
between the ground and excited potentials, effectively dividing the molecular medium into independent coherent
clusters ($J$ states), which experience gain at slightly different frequencies. Near threshold, only the most
populated clusters can oscillate, reducing the density available for gain to $1.5\!-\!5\%$ of the total density.
With these considerations, the threshold for $K_2$ medium in a cavity with $1\%$ loss is estimated to require molecular
density of $10^{13-14}\text{cm}^{-3}$ in a cell length of $5\!-\!10\text{cm}$, pumped by $\sim\!500$mW  at
50MHz repetition rate (see online material for detailed decoherence considerations and threshold estimation).

The selective and efficient dumping to a single state is important since it enables unique reconstruction of the
excited wave-packet dynamics from the emitted field. Specifically, spectral analysis of the emitted field fully
reflects the vibrational structure of the excited potential (including phase), as detailed in the online material.
We do not view the Raman oscillator as a new method to transfer efficiently population between ground levels,
since it is rather complicated, and since other well established, robust and efficient methods exist for population transfer between ground states, such as STIRAP \cite{RevModPhys.70.1003}.

To conclude, a coherent Raman oscillator that amplifies emission from
a coherently excited wave-packet, appears within experimental reach. If realized, this oscillator can open a window to
explore molecular coherent dynamics by amplifying the emitted signal per molecule
by several orders of magnitude. The unique effects of mode competition between different
coherent transfer possibilities in such an oscillator are of great interest.

This research was supported by the Israel Science Foundation (grants \#807/09 and \#46/14).

\clearpage
%

\begin{widetext}
\section{\label{sec:supplementary_material_crl} Supplementary Material} 
\subsection{\label{sec:single_pass_gain}Single-pass gain}
This section reviews the calculation of the single-pass gain in every iteration of the cavity simulation.
Sub-section \ref{subsec:single_molecule_field_emission}
describes how the microscopic dipole emission of each molecule is obtained from a
Schr\"{o}dinger calculation of the molecular dynamics;
sub-section \ref{subsec:macroscopic_field_emission} analyses the macroscopic field emission from the molecular ensemble
by a calculation of the number of excited molecules within the spatial mode volume of the oscillator;
sub-section \ref{subsec:pump_absorption_photon_budget} incorporates the effect of pump absorption
(``photon budget'') within the gain medium;
and sub-section \ref{subsec:optimal_intra_cavity_focusing} explores the optimal intra-cavity focusing for the highest single-pass gain with minimal pump depletion.

We analyze the oscillator under a low-gain assumption, where the medium is optically thin for the intra-cavity accumulated field; i.e.,
all molecules across the medium experience the same stimulating dump field (but not necessarily the same pump intensity, due to pump absorption).
This assumption is reasonable when the single-pass gain is $<\!15\!-\!20\%$, which is well satisfied by our configuration and simulation.
For the pump excitation however, propagation and depletion of the pump pulse are fully-included in the calculation for any optical thickness of the gain medium.

\subsubsection{Single-molecule field emission} \label{subsec:single_molecule_field_emission}
The oscillating electric dipole in the medium emits a temporal field $E^{em}(z,t)$ (under the dipole approximation)
\begin{small}
    \begin{align}
        \label{eq:far_field}
        E^{em}(z,t)
        & =
            \dfrac{1}{4\pi \varepsilon_0 c^2 z}
            \dfrac{\ud^2}{\ud t^2}
            P_M(t - z/c),
    \end{align}
\end{small}
where $P_M(t)$ is the temporal total (macroscopic) dipole moment of all molecules in the medium, and $z$ is the distance of the
observation point from the molecular medium. Equation \ref{eq:far_field} assumes para-axial emission from an isotropic gain-medium,
where the dipole direction $\vb{\hat{\mu}}$ is along the pump polarization and perpendicular to the
optical axis of the cavity $\vb{\hat{z}}$ (for a more general expression cf. \cite{BornWolf:1999:Book,jackson_classical_1999}).

In a large molecular ensemble, the total dipole is a sum over all single-molecule contributions $P_M(t)\!=\!\mathcal{N}P(t)$,
where $P(t)$ is the quantum average dipole of a single molecule and $\mathcal{N}$ is the effective number of molecules.
In this sub-section, we detail the calculation of $P(t)$ and its contribution to the emission,
whereas the calculation of $\mathcal{N}$ is detailed in sub-sections B and C. The quantum average dipole of a molecule is
\begin{small}
    \begin{align}
        \label{eq:dipole_moment}
        \begin{split}
            P(t)
            & =
                \Big[
                    e^{- i \omega t}
                    \mu_{eg}
                    \braket{ \psi_{g}(t) | \psi_{e}(t) }
                    +
                    \text{cc}
                \Big]
            \\
            &
            \equiv
                p(t) e^{-i \omega t}
                +
                \text{cc},
        \end{split}
    \end{align}
\end{small}
where $\ket{\psi_{g}(t)}, \ket{\psi_{e}(t)}$ are the vibrational wave-functions on the ground ($g$) and excited ($e$) electronic potentials and
$\mu_{eg}\!=\!\bar{\mu}_{ge}\!\equiv\!\mu$ is the electronic dipole moment between the potentials, assumed to be independent of the
inter-nuclear distance. The frequency $\omega\!=\!\omega_0\!-\!i\gamma$ ($\omega_0$ the optical frequency) can be slightly complex to represent
homogeneous decay of the macroscopic dipole on a time scale of $\tau_c\!=\!1/\gamma$. $\ket{\psi_{e}(t)}$
is viewed in a rotating frame at frequency $\omega_0$ and $p(t)\!\equiv\!\mu\braket{ \psi_{g}(t) | \psi_{e}(t) }$ is the single-molecule dipole amplitude,
which reflects the time-dependent Franck-Condon overlap between the ground and excited vibrational wave-functions.
According to \eqref{eq:far_field}, the dipole emission from each molecule is proportional to the $2^{\text{nd}}$ time-derivative of $P(t)$
\begin{small}
    \begin{align}
        \label{eq:dipole_moment_2nd_derivative_classic}
        \begin{split}
            \ddot{P}(t)
            &
            =
                 \left[
                    - \omega^2 p(t) - 2 i \omega \dot{p}(t)+\ddot{p}(t)
                \right]
                e^{- i \omega t}
                +
                \text{cc}.
        \end{split}
    \end{align}
\end{small}
To express the emitted field, the dipole amplitude $p(t)$ and its time derivatives need to be calculated.
For this purpose, we use the split-operator method \cite{0034-4885-58-4-001} to numerically solve the time-dependent Schr\"{o}dinger equation for the vibrational wave-packets on the coupled
electronic potentials within the rotating-wave approximation (RWA).
\begin{small}
    \begin{align}
        \label{eq:schrodinger_equation}
        \begin{split}
            i \frac{\ud}{\ud t}
            \begin{pmatrix}
                \psi_{g}
                \\
                \psi_{e}
            \end{pmatrix}
            &\!=\!
                \begin{pmatrix}
                    T\!+\!U_{g}(R)    &   - \Omega^{\ast}(t)
                    \\
                    - \Omega(t)    &   T\!+\!U_{e}(R)\!-\! \omega_0
                \end{pmatrix}\!
                \begin{pmatrix}
                    \psi_{g}
                    \\
                    \psi_{e}
                \end{pmatrix}
        \end{split},
    \end{align}
\end{small}
where $R$ is the inter-nucleic distance, $T\!=\!\begin{small} -\frac{\hbar}{2M}\frac{\partial^2}{\partial R^2}\end{small}$
is the kinetic energy operator of the nuclei ($M$ the reduced mass), $U_{g,e}(R)$ is the ground/excited electronic potential (in units of frequency) and $\Omega(t)=\mu E(t)/\hbar$
is the slow-varying amplitude of the accumulated intra-cavity field (Rabi frequency).

Using the Schr\"{o}dinger equation \ref{eq:schrodinger_equation} we find
\begin{small}
 \begin{align}
    \label{eq:dipole_moment_1st_dertivative}
    \begin{split}
        -2i
        \omega
        \dfrac{\dot{p}(t)}{\mu}
        & =
            \bra{ \psi_{g}}
                2\omega\omega_0
                \!
                -
                \!
                2\omega U
            \ket{\psi_{e} }
            \!
            +
            \!
            2\omega\Omega(t)
            \left(
                \braket{ \psi_{e}|\psi_{e} }
                \!
                -
                \!
                \braket{ \psi_{g}|\psi_{g} }
            \right)
            \\
         \dfrac{\ddot{p}(t)}{\mu}
         & =
            \bra{ \psi_{g}}
                -U^2
                \!
                +
                \!
                2\omega_0 U
                \!
                -
                \!
                \omega_0^2
                \!
                -
                \!
                \left[T,U\right]
            \ket{\psi_{e} }
            +
            \Omega(t)
            \left(
                \bra{ \psi_{e}}
                    U
                    \!
                    -
                    \!
                    \omega_0
                \ket{\psi_{e} }
                \!
                -
                \!
                \bra{ \psi_{g}}
                    U
                    \!
                    -
                    \!
                    \omega_0
                \ket{\psi_{g} }
            \right)
        \\
        &
            \qquad
            \qquad
            +
            i\dot{\Omega}(t)
            \left(
                \braket{ \psi_{e}|\psi_{e} }
                \!
                -
                \!
                \braket{ \psi_{g}|\psi_{g} }
            \right)
            \!
            -
            \!
            2i\Omega(t)
            \text{Im}
            \left[
                \Omega(t)\braket{ \psi_{e}|\psi_{g} }
            \right],
    \end{split}
\end{align}
\end{small}
where $U$ is the electronic energy difference ($U\!\equiv\!U_{e}(R)\!-\! U_{g}(R)$).
Note that as long as the field (Rabi frequency) and its time derivative are small compared to the optical frequency ($\Omega^2, \dot{\Omega}\!\ll\!\omega\Omega$)
(the inherent RWA assumption), the last row of Eq. \ref{eq:dipole_moment_1st_dertivative} can be neglected.

Combining \eqref{eq:dipole_moment_1st_dertivative} into (\ref{eq:dipole_moment_2nd_derivative_classic}),
we obtain the $2^{\text{nd}}$ derivative of the single-molecule dipole (the microscopic emitted field)
\begin{small}
    \begin{align}
        \label{eq:single_molecule_em_field}
        \begin{split}
            \frac{\ddot{P}(t)}{\mu}
            =
                -\Big[
                    &
                    \bra{ \psi_{g}}
                        \left(
                            U-i\gamma
                        \right)^2
                        \!
                        +
                        \!
                        \left[
                            T,U
                        \right]
                    \ket{\psi_{e} }
                +
                \Omega(t)
                \Big(
                    \bra{ \psi_{e}}
                        U
                        \!
                        -
                        \!
                        i\gamma
                        \!
                        +
                        \!
                        \omega
                    \ket{\psi_{e} }
                -
                    \bra{ \psi_{g}}
                        U
                        \!
                        -
                        \!
                        i\gamma
                        \!
                        +
                        \!
                        \omega
                    \ket{\psi_{g} }
                \Big)
                \Big]
                e^{- i \omega t}
                +
                \text{cc},
        \end{split}
    \end{align}
\end{small}
where terms from the last row of \eqref{eq:dipole_moment_1st_dertivative} were neglected in accordance with the RWA.
Thus, by calculating the vibrational wave-packets in time $\ket{\psi_{g}(t)}, \ket{\psi_{e}(t)}$
after each pump pulse, the emitted field in every cavity round trip can be obtained.

Equation \ref{eq:single_molecule_em_field} indicates \emph{\textbf{two contributions to the emitted field (gain)}}:
The second term represents the ``standard'' stimulated emission due to the (time-dependent) \emph{population inversion}
between the ground and excited states, proportional to the inducing field $\Omega(t)$. 
The first term however, is unique, representing \emph{dipole emission from a coherent superposition of ground and excited wave-packets}, 
which does not require an inducing field. As we show hereon, the emission is dominated by the coherent dipole term most of the time after the pump pulse, 
while standard stimulation is important only to seed the emission at the very early stage. Equation \ref{eq:single_molecule_em_field} thus forms a coherent wave-packet generalization
of the rate equation for the gain in a standard laser \cite{siegman86}.
An atomic version of Eq. \eqref{eq:dipole_moment_1st_dertivative} appears in \cite{laser_physics_and_applications_vol_1}.

The single-molecule dynamics is calculated in two stages. First, we calculate the excited wave-packet immediately after
the (short) pump pulse by solving the Schr\"{o}dinger equation \eqref{eq:schrodinger_equation} with all population initially in the vibrational
ground state of the ground electronic potential, and the pre-defined pump pulse as the coupling field.
Then, this excited wave-packet serves as the initial state for calculating the excited-target
dynamics, which is induced by the accumulated intra-cavity dump field, providing the time-dependent dipole of a single molecule (microscopic)
and single-molecule emitted field. Note that although the target state resides in the ground electronic potential, the residual un-pumped
population in the ground vibration $\ket{\nu \!=\!0}$ does not affect the emission and we can treat the target potential as empty.
This is because the target wave-packet occupies much higher vibrational states of the ground potential (near $\ket{\nu \!=\!23}$ in $K_2$),
which are thermally empty, and the cavity mirrors resonate only the Raman shifted emission, but not the pump light.

This two-stage calculation of "first pump, then dump" is justified since the emission
of the dump field appears several picoseconds after the pump excitation, and is well separated in time from the exciting pump.
We verified that no early-stage dynamics was ignored by this two-stage separation with an additional precise simulation of all three potentials
(ground-excited-target) coupled simultaneously by both the pump and the dump fields
(similar to the solution in \cite{PhysRevLett.98.113004}), which showed no notable difference from the two-stage simulation.

\subsubsection{\label{subsec:macroscopic_field_emission} Macroscopic field emission}
When summing the microscopic emissions of single molecules to calculate the macroscopic emitted field, it is necessary to account for macroscopic
decoherence due to collisions and other broadening mechanisms. We therefore add a phenomenological temporal decay of
the average dipole $P(t)\!=\!p(t)e^{-i\omega_0 t}e^{-\gamma t}$ on a time scale $\tau_c\!=\!1/\gamma$ (of exponential profile, assuming pressure
broadening as the major decoherence mechanism), which renders the frequency $\omega$ slightly complex $\omega\!=\!\omega_0\!+\!i\gamma$.

The macroscopic emitted field (single-pass gain) can now be calculated with the following considerations:
a) The spatial mode of the accumulated field is Gaussian, which corresponds to the $\text{TEM}_{00}$ mode
   of the laser cavity and is matched to the pump spatial mode.
b) The spatial mode of the emitted field is inherently that of the inducing field.
   This is an immediate result of stimulated emission with linear dipole response, i.e. the local dipole emission is proportional to the local inducing
   field.
c) As a direct result of Fresnel diffraction, the macroscopic field on the optical axis $E_{M}^{em}(z)$ at a large enough distance from the beam waist
   ($z \gg z_R$, the Rayleigh range), is just the coherent in-phase sum of all the microscopic dipole contributions
\begin{small}
    \begin{align}
        \label{eq:far_field_em}
        E_{M}^{em}(z)=\mathcal{N} E_{\text{single}}^{em}(z)\sim\mathcal{N}\frac{\ddot{P}}{z},
    \end{align}
\end{small}
where $\mathcal{N}$ is the number of molecules within the effective illuminated volume.

To provide a closed expression for the macroscopic emitted field,
we place the molecular medium at the waist of the cavity mode and calculate the emitted field at a distance $z\!=\!20z_R$,
well within the far-field range. By diffraction, this on-axis field is the center value of the emitted Gaussian spatial mode,
which is identical to the inducing Gaussian mode. Using Gaussian optics we propagate the emitted field back to the waist
\begin{small}
\begin{align}
    \label{eq:near_field_em}
    E_{M}^{em}(0)=\mathcal{N} E_{\text{single}}^{em}(z)\frac{i z}{z_R}\sim \mathcal{N}\frac{i \ddot{P}}{z_R},
\end{align}
\end{small}
where the factor $i$ reflects the Gouy phase shift between the far field and the focus. This emitted field is then added to
the previous inducing field (with appropriate loss and dispersion) to obtain the inducing intra-cavity field for the next iteration.
The macroscopic single-pass gain is thus expressed based on the microscopic emission extracted from the simulation.

\subsubsection{\label{subsec:pump_absorption_photon_budget} Pump absorption - ``photon budget''}
We now calculate the effective total number of molecules $\mathcal{N}$, relating the microscopic field emission to the macroscopic single-pass gain,
which requires to estimate the effective volume of the gain medium. We assume that the gain medium is
located at the focus of the pump beam with area $A\!=\!\pi w_0^2\!=\!z_R\lambda$ ($w_0$ the waist radius, $z_R$ the Rayleigh range) and
the length of the medium is matched to the depth of focus $L\!=\!2z_R$.
If we also assume equal pumping across the illuminated volume (thin medium for the pump pulses),
the effective volume would just be the focal volume $V_{\text{eff}}^0\!\approx\!2z_R^2\lambda$. Here however, a thin medium approximation may be over-simplified,
as absorption of the pump pulses during propagation may be substantial.
Thus, the effective medium length can be shorter due to pump depletion, and careful management of the
pump ``photons budget'' is important for calculation of the actual effective volume.

If the excited population of a single molecule is not high (no saturation), the excited population $|\psi_{e}|^2$
of a molecule at position $z$ is proportional to the local pump-pulse energy density: $|\psi_{e}|^2(z)\!=\!\beta \varepsilon_{p}(z)/A$, where $\varepsilon_p(z)$ is the
energy of the pulse at position $z$ and $A$ is the cross section of the beam $A\!=\!\pi w_0^2\!=\!z_R\lambda$.
$\beta$ is a power-independent absorption factor, which can be determined numerically by calculating the excited population at $z\!=\!0$
for various $\varepsilon_p(0)$, all with the same pump pulse shape (verifying that the absorption is linear with no saturation).

The number of absorbed pump photons $\ud N_{\text{ph}}$ in a length element $\ud z$ is equal to the number
of excited molecules according to $\ud N_{\text{ph}}(z)\!=\!\ud N_{\text{ex}}(z)\!=\!|\psi_{e}|^2(z)\rho A \ud z$
($\rho$ is the molecular density). Thus, the energy absorbed in a depth element $\ud z$ is
\begin{align}
    \ud \varepsilon_p(z)
    & =
        - \hbar \omega \ud N_{\text{ph}}
    =
        - \hbar \omega \beta \rho \varepsilon_p(z) \ud z,
\end{align}
which leads to
\begin{align}
    \dfrac{\ud \varepsilon_{p}(z)}{\ud z}
    & =
        - \hbar \omega \beta \rho \varepsilon_{p}(z)
    \equiv
        - \alpha \varepsilon_{p}(z),\ \ \ \ \  \varepsilon_{p}(z)= \varepsilon_{p}(0)e^{- \alpha z} .
\end{align}
The number of excited molecules $\ud N_{\text{ex}}(z)$ for a depth
$\ud z$ is $\ud N_{\text{ex}}(z)\!=\!\beta \rho \varepsilon_p(z) \ud z$,
indicating that the total number of excited molecules after traversing a medium of length $L\!=\!2 z_R$ is
\begin{align}
    \label{eq:no_of_excited_mol}
    N_{\text{ex}}
    & =
        \int_{0}^{L}
            \beta \rho \varepsilon_{p}(z) \ud z
    =
        N_{\text{ph}}\Big( 1 - e^{- \alpha L} \Big),
\end{align}
where $N_{\text{ph}} = \varepsilon_{p}(0)/ \hbar \omega$ is the total number of photons in the input pump pulse.
Thus, the effective length of the medium is normalized according to the pump depletion $L_{\text{eff}}\!=\!(1\!-\!e^{-\alpha L})/\alpha$, and the effective number of molecules is
\begin{align}
    \label{eq:effective_no_of_mol}
    \mathcal{N}
        &\!=\!\rho A L_{\text{eff}}\!=\!\rho z_R\lambda \frac{1-e^{-\alpha 2 z_R}}{\alpha}\!\rightarrow\!2\rho z_R^2\lambda,
\end{align}
where for low pump depletion $L_{\text{eff}}\!\rightarrow\!2z_R$.

\subsubsection{\label{subsec:optimal_intra_cavity_focusing} Optimal intra-cavity focusing}
We can now estimate the optimal intra-cavity focusing condition by exploring the effect of the Rayleigh range $z_R$ on the single-pass gain and on the pump depletion. 
Combining equations \ref{eq:near_field_em} and \ref{eq:effective_no_of_mol} we obtain
\begin{align}
    \label{eq:near_field_em_final_1}
    E_{M}^{em}(0)\sim\mathcal{N}\frac{\ddot{P}}{z_R}\!=\!\ddot{P}\rho \lambda \frac{1-e^{-\alpha 2 z_R}}{\alpha}\!\rightarrow\!2\ddot{P}\rho z_R\lambda.
\end{align}
If we assume, similar to a standard laser, that the emitted field during the amplification stage
(when population of the target state is low $|\psi_{e}|^2\!-\!|\psi_{g}|^2\!\approx\!|\psi_{e}|^2$)
is proportional to the product of the intra-cavity inducing field and the excited population
$\ddot{P}\!\sim\!E|\psi_{e}|^2\!\sim\!E\varepsilon_{p}/A\!=\!E\varepsilon_{p}/(\lambda z_R)$, the macroscopic emitted field gain (the ratio of emitted field to intra-cavity inducing field) becomes
\begin{align}
    \label{eq:near_field_em_final_2}
    \frac{E_{M}^{em}(0)}{E}\sim\frac{\varepsilon_{p}}{z_R}\rho \frac{1-e^{-\alpha 2 z_R}}{\alpha}\!\rightarrow\!2\varepsilon_{p}\rho.
\end{align}

Consequently, as long as the pump depletion is low (thin medium for pump) and the excited state is not saturated,
the macroscopic small-signal gain is invariant to focusing and depends only on the energy-density product of pump-pulse energy and molecular density.
This invariance is limited either by substantial pump depletion for very low focusing or by saturation of the molecular excited state for very tight focusing.
Table \ref{table:gain} shows the simulated single-pass gain for various beam radii $w_0$ with
pump pulse-energy of 15nJ and molecular density of $3\cdot10^{12}/cm^3$.
The optimal focusing in this case is a beam radius of $w_0\sim\!150\mu m$, where a relatively high intensity gain of $\sim\!17$\% is obtained with only $\sim\!5$\%
pump absorption.
This low pump absorption enables reuse of the pump energy and lowering of the required pump laser power even more,
by adding an enhancement cavity for the pump, whose free-spectral range is matched to the
repetition rate of the pulses and input transmission corresponds to the pump absorption in the sample.
Thus, with a low-finesse cavity for the pump, the required $15$nJ of pump energy,
can be easily supplied by a $1$nJ source ($100$mW average power at a $100$MHz repetition rate),
which is conveniently achievable with standard mode-locked oscillators. 

\newcolumntype{d}[1]{D{.}{.}{#1}}
\begin{table}
    \begin{minipage}{\textwidth}
        \caption{Calculated single-pass intensity gain, for various beam focal radii $w_0$, given a pump pulse-energy of $\varepsilon_p\!=\!15 \text{nJ}$,
                 and molecular density $\rho\! =\! 3 \cdot 10^{12}$ $\text{Molecules}/\text{cm}^3$.
                 The gain is evaluated at the early stage of amplification, where the population transfer to the target is negligible.
                 $w_0$ is the beam radius at the waist, $z_R$ is the Rayleigh range, $|\psi_{e}|^2$ is the population of the excited state at the input to the
                 medium ($z\!=\!0$) immediately after the pump pulse.
                 Gain is diminished for $w_0 > 400 \mu m$ by significant pump absorption, and for $w_0 < 125\mu m$, by saturation of the excited population $|\psi_{e}|^2$.
                 }
        \begin{tabular}{d{7.4}d{7.7}d{7.7}d{7.7}d{7.7}}
            \hline\noalign{\smallskip}
            \multicolumn{1}{c}{$w_0$ }
                         &  \multicolumn{1}{c}{$z_R$}
                                        &  \multicolumn{1}{c}{$|\psi_{e}|^2$ }
                                                         &  \multicolumn{1}{c}{Gain}
                                                                         &  \multicolumn{1}{c}{Absorption}
            \\
            \multicolumn{1}{c}{[$\mu m$]}
                         &  \multicolumn{1}{c}{[cm]}
                                        &  \multicolumn{1}{c}{[\%]}
                                                         &  \multicolumn{1}{c}{[\%]}
                                                                         &  \multicolumn{1}{c}{[\%]}
            \\
            \hline\noalign{\smallskip}
            500         &  77.91        &   2.39         &  13.81        &   48.47
            \\
            400         &  49.87        &   4.21         &  16.14        &   30.75
            \\
            300         &  28.05        &   9.22         &  17.66        &   14.49
            \\
            200         &  12.47        &  11.86         &  17.76        &   11.05
            \\
            175         &   9.54        &  15.76         &  17.62        &    7.96
            \\
            150         &   7.01        &  21.82         &  17.12        &    5.28
            \\
            125         &   4.87        &  31.72         &  15.93        &    3.06
            \\
            100         &   3.12        &  48.36         &  13.33        &    1.37
            \\
            75          &   1.75        &  71.33         &   7.50        &    0.31
            \\
            \hline\noalign{\smallskip}
            \hspace{4cm} & \hspace{4cm} & \hspace{4cm}   & \hspace{4cm}  & \\
        \end{tabular}
        \label{table:gain}
    \end{minipage}
\end{table}

\subsection{\label{sec:threshold_effects_of_temperature_and_pressure} Threshold - Effects of Temperature and Pressure}
\label{sec:threshold_ana_available_coherence_time}

The calculated gain of table \ref{table:gain} predicts reasonably high gain at rather low pump energies. 
A preliminary estimation of the oscillation threshold can now be obtained from the calculated gain and the cavity losses: Given the pump pulse energy,
one can estimate from the energy-density product the necessary molecular density to overcome the cavity losses, 
and select a convenient focusing parameter (Rayleigh range) to satisfy low pump-depletion and low excited-state saturation. 
This threshold estimation however is only preliminary since the gain will vary according to the actually available coherence time 
(so far fixed at $25$ps for computational reasons only), and since the calculation did not take into account coherent gain dynamics during the 
oscillation buildup (unique to this coherent Raman oscillator and does not appear in standard lasers), 
or the inhomogeneous thermal distribution of the molecular population, which divides the population to independent coherent clusters and reduces 
the available population in a single coherent cluster. These effects are discussed in the following subsections.

\subsubsection{\label{subsec:coherent_gain_dynamics_during_cavity_buildup} Coherent gain dynamics during cavity field build-up}
The small signal gain of Eq. \eqref{eq:near_field_em_final_2} and the related energy-density product are similar to a standard laser,
and provide simple and convenient relations for a basic assessment of the lasing threshold.
Yet, Eq. \eqref{eq:near_field_em_final_2} enfolds an inherent assumption that is not necessarily correct - it assumes that during the amplification stage,
the gain is local in time (the emitted field at time $t$ is proportional to the field $E(t)$ at the same time).
While this is correct for a standard laser, where the emitted field corresponds only to the stimulated emission term
$E\left(\left|\psi_e\right|^2\!-\!\left|\psi_g\right|^2 \right)$, our coherent Raman oscillator
\eqref{eq:single_molecule_em_field} includes an additional gain term of coherent dipole emission $\bra{\psi_{g}} U^2\!+\![T,U] \ket{\psi_{e}}$,
which dominates the emitted field well before any significant population is dumped to the target state.
The dependence of the coherent dipole emission term on the inducing field $E(t)$ is not local in time,
but rather global - the emission at time $t$ depends on the dipole at $t$, which in turn depends the entire field evolution since the pump pulse up to $t$.

The evolution of the gain during cavity build-up is therefore more complicated than that of the standard laser,
as was indeed observed in our simulation and is shown in figure \ref{fig:gain_plot}.
The gain is not constant during build-up, but tends to decay well before any
significant population is dumped to the target state (the final populations of the target
and excited states after the end of the dump-pulse window are also shown for comparison).
This global gain dynamics also tends to shape the emitted pulse in time during the amplification stage,
since parts of the intra-cavity dump pulse at later time enjoy higher gain than earlier parts (a larger coherent
dipole could develop by the later time).

For example, our simulation showed for a gain medium of $K_2$ molecules in a cavity with 3\% linear losses that is pumped 10 nJ pulses,
the actual threshold density (including the coherent gain reduction during buildup) was $\sim 3 \cdot 10^{12}/\text{cm}^3$ at an interaction length of $2z_R\!\sim\!16\text{cm}$, 
somewhat higher than expected based on the low-signal gain of section \ref{subsec:optimal_intra_cavity_focusing}.
The threshold density for $Rb_2$ dimers was very similar, whereas for $Li_2$ it was
slightly higher ($\sim10^{13} \text{molecules}/\text{cm}^3$)
due to the lower transition dipole-moment of $Li_2$ compared to $K_2$.

\begin{figure}
    \begin{minipage}[t]{0.45\linewidth}
      \includegraphics[width=\linewidth,height=7cm]{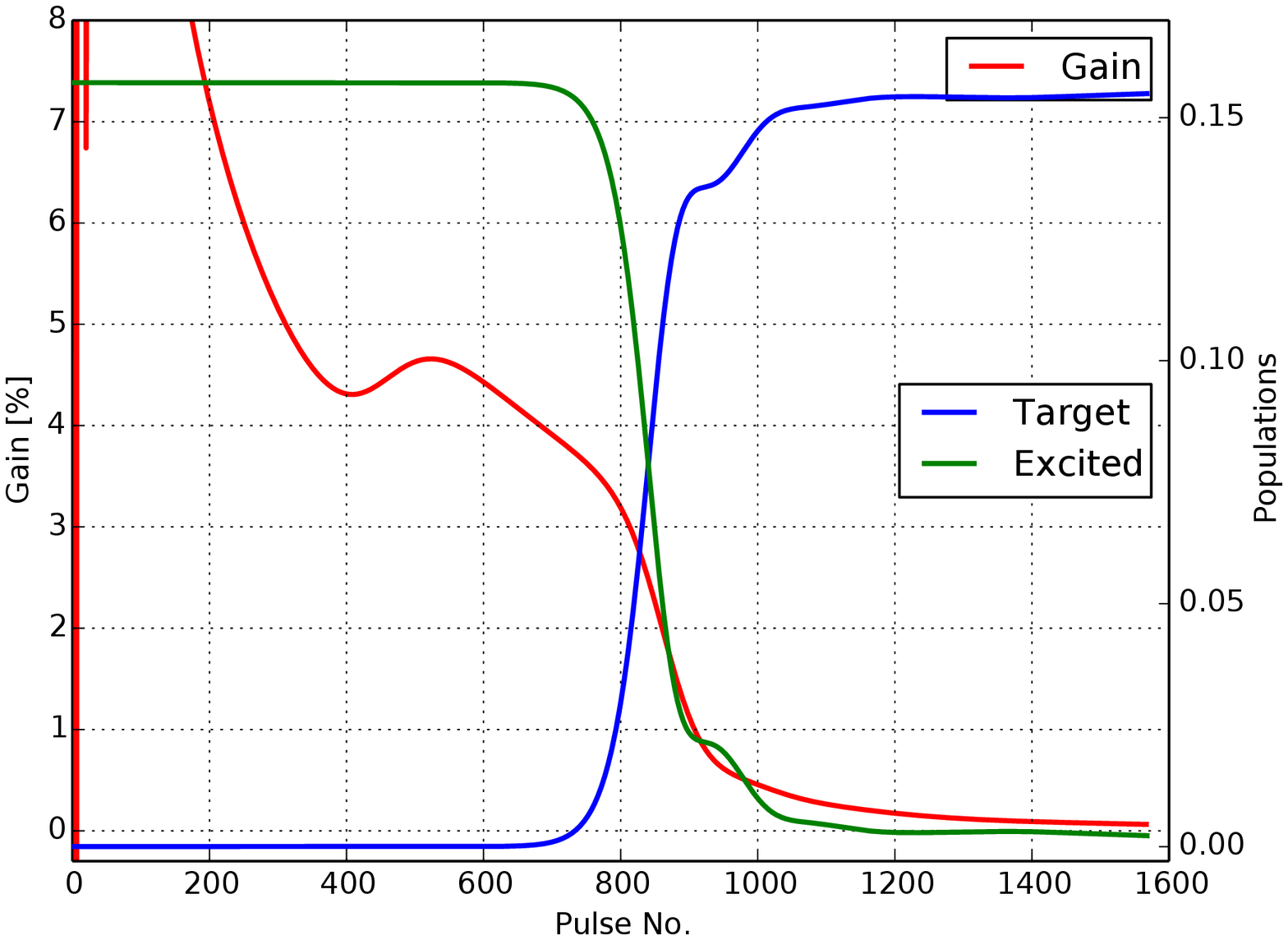}
      \caption
      {
          Coherent gain dynamics during cavity buildup: The red curve shows the intensity-gain ($K_2$) 
          of the accumulated intra-cavity field as a function of time during cavity build-up, 
          scaled in cavity round-trips (pump pulses) since the beginning of the simulation.
          In addition, the final populations at the end of the 50ps simulation window, are shown for the excited state (green) and the target state (blue), 
          demonstrating the complete coherent population transfer near pump-pulse 850.
          Even well before the transfer takes place, the gain per pulse is not constant and tends to diminish during buildup.
      }
      \label{fig:gain_plot}
    \end{minipage}
    \begin{minipage}[t]{0.025\linewidth}
    \hspace{0.2\linewidth}
    \end{minipage}
    \begin{minipage}[t]{0.45\linewidth}
          \includegraphics[width=\linewidth,height=7cm]{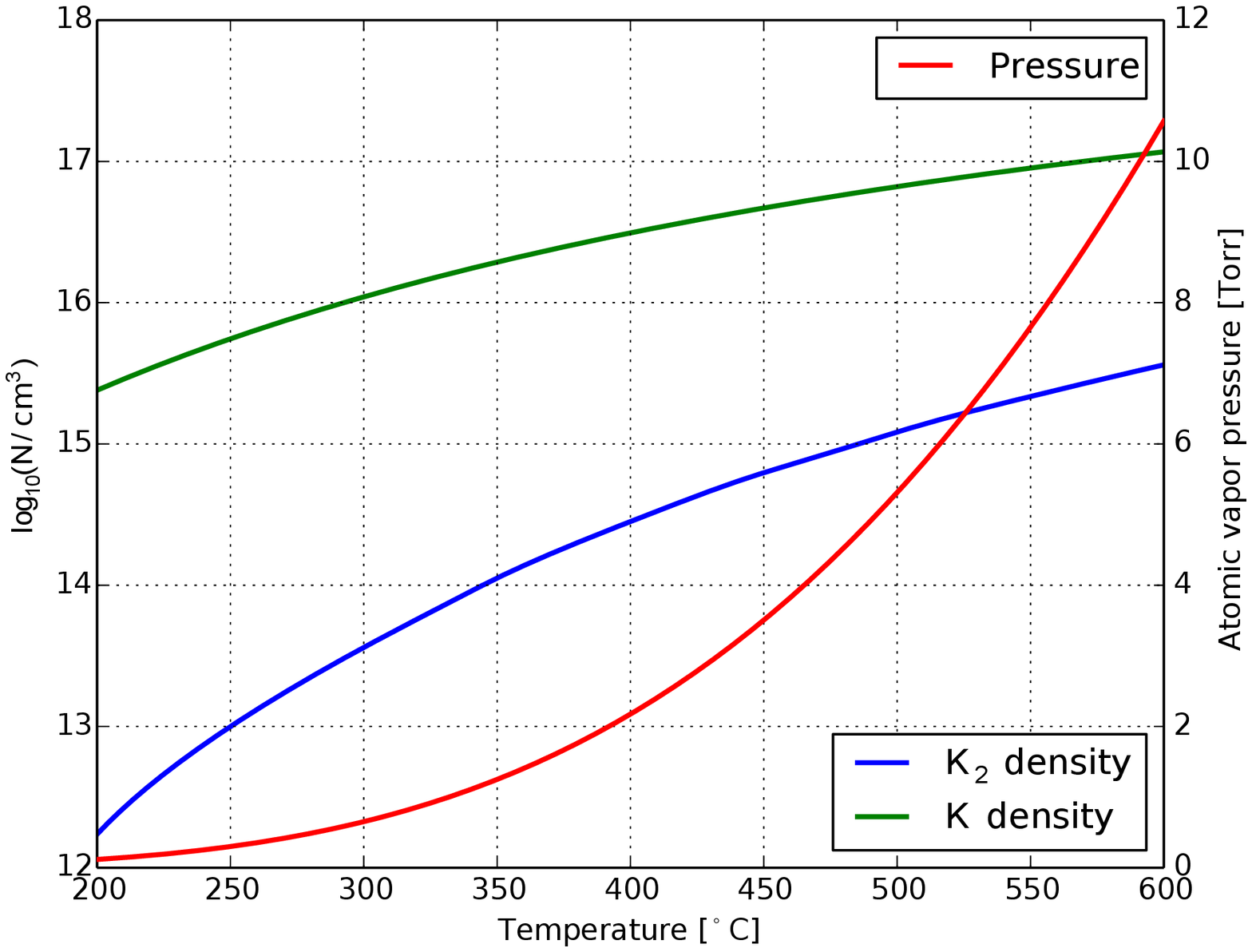}
          \caption
          {
              Saturated vapor densities in thermal equilibrium for $K_2$ dimer (blue) and $K$ atoms (green)
              as a function of temperature.
              Data was taken from \cite{lapp_harris_1965,doi:10.1179/cmq.1984.23.3.309,0022-3700-15-18-005,1980JETPL..31..554A}.
              The vapor pressure of the atomic $K$ gas (red) is also shown.
              The dimer density of $10^{12}/\text{cm}^3$ can be achieved already at $\sim\!200^{\circ}C$ ($\sim\!470^{\circ}K$),
              with atomic vapor pressure lower than 1Torr.
          }
        \label{fig:dimer_density}
    \end{minipage}
\end{figure}

\subsubsection{\label{sec:homogeneous_broadening} Homogeneous broadening}
Due to the above coherent gain dynamics, the oscillation threshold (as well as other properties of the emitted field)
depends strongly on the available coherence time.
The decoherence window in our simulation was assumed to be $T_2 \approx25\text{ps}$ (mainly due to computational limitations),
which is very conservative for the relevant temperature and pressure conditions.
In this sub-section we estimate the coherence time by considering the collisional pressure broadening that accompanies
the high density in a realistic chamber of hot alkali vapor.
Discussion of the effects of inhomogeneous broadening is deferred to the \hyperref[sec:inhomogeneous_broadening]{next sub-section}.

The threshold densities in the range of $10^{12\!-\!13} \text{ molecules}/\text{cm}^3$ can be achieved in a vapor cell at a
temperature range of $500^{\circ} - 550^{\circ}$K, as shown in figure \ref{fig:dimer_density}
(based on \cite{doi:10.1179/cmq.1984.23.3.309,0022-3700-15-18-005,1980JETPL..31..554A,PhysRev.55.1267}).
Since the dimer molecules constitute only 0.2-1\% of the density at these temperatures  \cite{1980JETPL..31..554A,lapp_harris_1965},
the major broadening source for $T_2$ is collisions with the surrounding atoms of density $10^{14-15} \text{ atoms}/\text{cm}^3$,
indicating that the vapor pressure in the cell will be at most 1-5 Torr \cite{doi:10.1179/cmq.1984.23.3.309}.
The exact pressure broadening for $K-K_2$ collisions is not known to us, but the atomic pressure broadening value
($K - K$ collisions) for the $D_1$ line is $\sim\!2\text{ GHz}/\text{Torr}$ \cite{siegman86}.
This is an anomalously high value, since it is affected by resonant dipole-dipole collisions, which have a large cross-section.
For non-resonant collisions, typical broadening values are of a few $10-100 \text{MHz}/\text{Torr}$ \cite{PhysRev.55.1267,PhysRev.136.A1233,Pitz2012387}.
For $K_2$ the molecular transitions are far detuned from the atomic lines, and therefore $K-K_2$ collisions are expected to be predominantly non-resonant.
Yet, even if we adopt the severe atomic broadening value, the collisional coherence is expected to be $>100 \text {ps}$,
which is more than sufficient for the coherent Raman oscillator. In fact, it is even likely that the collision-rate will need to
be increased with a buffer gas (He or $N_2$) in order to expedite the transfer of population from
the target level back to the ground level and empty the target level before the next pump pulse.

Although we could not simulate longer coherence times directly (due to limited calculation power), we can estimate the reduction of threshold
due to a longer
coherence based on the well-known coherent dynamics of a two-level system.
When oscillating above threshold, the Raman oscillator generates coherent molecular ``$\pi-$pulses'', which dump the entire
excited wave-packet population
to the target state in a single coherent stroke of a coherent pulse-train with a period that matches the vibrational time of the excited wave-packet. Since the energy required for a $\pi-$pulse is inversely proportional to the pulse duration, the oscillator near threshold tends to
generate the longest ``$\pi-$pulse'' allowed by decoherence.
Thus, \emph{the threshold will be inversely proportional to the available coherence time},
which indicates that if the coherence time will indeed be $~\!100$ps, the density threshold (or pump energy) can be reduced by a
factor of 4 compared to the simulation. This conclusion however is subject to considerations of inhomogeneous broadening, which were so far ignored and are discussed in the next sub-section.

\subsubsection{\label{sec:inhomogeneous_broadening} Inhomogeneous broadening - Doppler and rotational degrees of freedom}
Inhomogeneous broadening mechanisms, such as Doppler or the thermal broadening from the
distribution of rotational states should also be considered. Inhomogeneous broadening
does not affect the coherence time of the individual molecule, which remains limited by collisional broadening,
but divides the molecular ensemble to many independent 'coherent clusters' (of different velocity / rotational states)
that emit at different carrier frequencies and do not compete among themselves for the gain resources.
Inhomogeneous broadening therefore reduces the available gain due to the reduced available density within a single coherent cluster,
indicating that higher molecular densities and higher pumping energies will be required to cross threshold.
Inhomogeneous broadening however will not affect the bandwidth (or coherence) of the oscillation in a single coherent cluster,
which is specifically relevant near threshold, where only a few coherent clusters (or just one) will actually oscillate
(those with the highest gain).
When the gain is increased further
above threshold, more coherent clusters will begin to oscillate independently (multi-mode operation), which will degrade the coherence of
the emission.

The Doppler broadening of the relevant optical emission frequencies for alkali dimers at $400-700^\circ K$
is of order 0.5-1GHz (corresponding to $T_2^*\!\sim\! 1 \! - \! 2$ns), which poses no limitation for
observation of coherent dynamics on timescales of 100ps (or more). Doppler can therefore be safely ignored.
The broadening due to the thermal distribution of rotational levels is more important. Using a rigid rotor model, we can estimate the
rotational broadening, based on the difference of rotational constants between the excited vibrational states and
the target vibration. The rotational energy for a state of angular momentum $J$ and vibrational level $\nu$ is
\begin{align}
    \label{eq:rigid_rotor_approximation}
    K^{J}_{\nu}
    & =
        \dfrac{\hbar^2 J(J+1)}{2I_{\nu}}
    ,
    &
    I_{\nu}
    & \equiv
        \mu \braket{R^2}_{\nu}
    =
        \mu \bra{\nu} \hat{R}^2 \ket{\nu},
\end{align}
where $I_{\nu}$ is the molecular moment of inertia at vibration $\nu$, dictated by the average distance between the nuclei $\braket{R^2}_{\nu}$. At a temperature of $500^{\circ}K$, the rotational states of the ground vibration $\nu_g=0$ are thermally excited up to $J \sim 100$, centered around $J \sim 55$ with a maximal fractional population of $\sim1.5\%$ in a single $J$ state (see figure  \ref{fig:K2_rotational_broadening}). Since the rotational state of the molecule remains unchanged in a Raman transition where both the pump and the inducing field have the same polarization, each of those rotations interacts with the light-field independently. Due to the difference in moment of inertia between the center excited vibrational mode $\nu_e$ and the target mode $\nu_f$, the emission frequency of each rotational $J$ state is slightly shifted by
\begin{align}
    \label{eq:rotational_broadening}
    \Delta^{J}_{\text{rot}}
    & =
        K^{J}_{f} - K^{J}_{e}
    =
        \dfrac{\hbar^2 J(J+1)}{2 \mu}
        \Bigg[
            \dfrac{1}{\braket{\nu_f| \hat{R}^2 | \nu_f} }
            -
            \dfrac{1}{\braket{\nu_e| \hat{R}^2 | \nu_e} }
        \Bigg],
\end{align}
as shown in figure \ref{fig:K2_rotational_broadening} (blue curve).

\begin{figure}
    \begin{minipage}[t]{0.45\linewidth}
      \includegraphics[width=\linewidth, trim=19cm 0 0 0, clip=true]{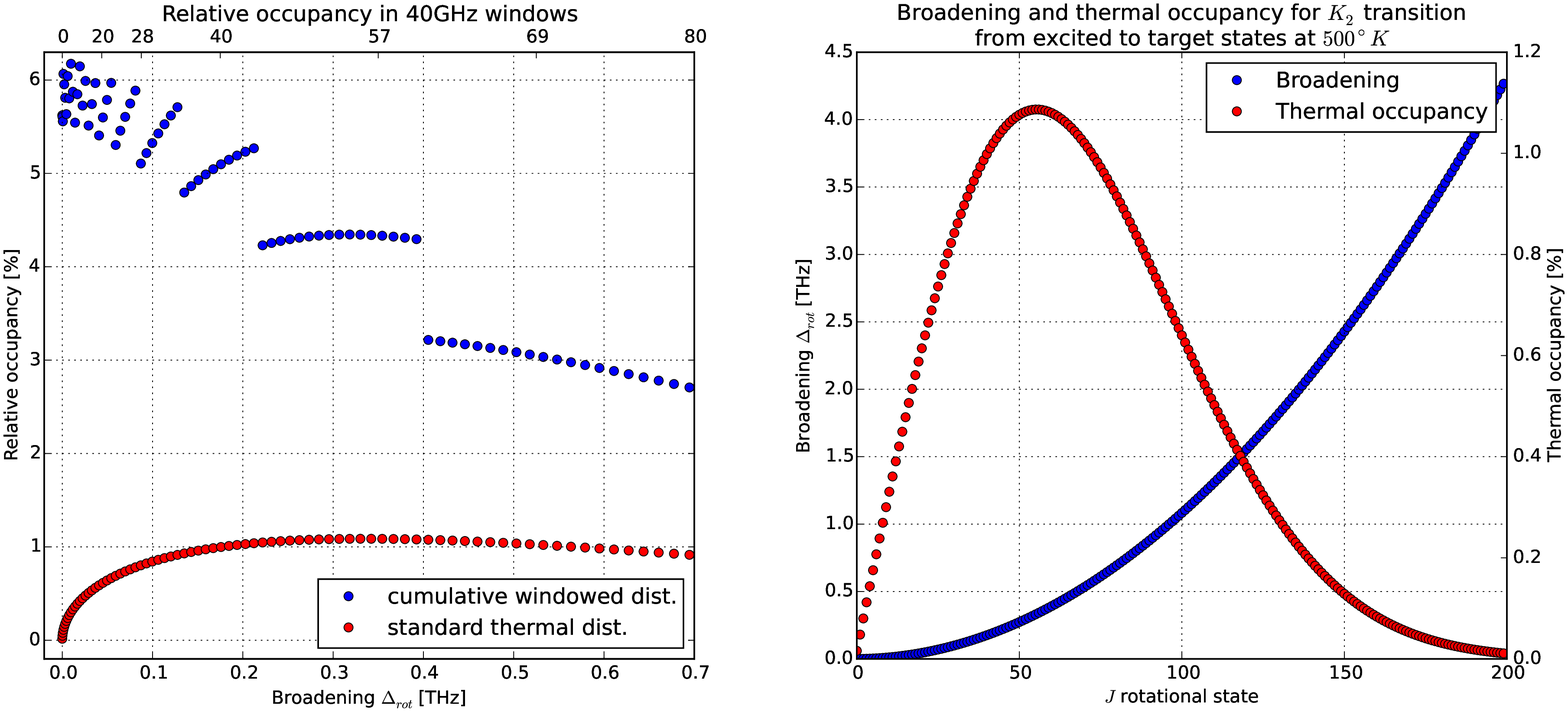}
      \caption
      {
          The rotational shift of transitions between excited vibrational mode $\nu_e = 10$ to target mode $\nu_f = 23$ in $K_2$ for various rotational states $J\!=\!0\!-\!200$ (blue).
          The relative population of each rotational state at $500^{\circ}K$ is shown in red
          $\rho(J)\! \propto \! (2J+1) \exp^{- E_{J} / k_B T}/ \sum_{J} N_J$, where $E_J$ is taken as a rigid rotor
          from Eq. \eqref{eq:rigid_rotor_approximation}
          with respect to the average position $\braket{\nu_g| \hat{R}^2 | \nu_g}$ in the original ground state $\nu_g = 0$.
      }
      \label{fig:K2_rotational_accumulation}
    \end{minipage}
    \begin{minipage}[t]{0.025\linewidth}
        \hspace{0.2\linewidth}
    \end{minipage}
    \begin{minipage}[t]{0.45\linewidth}
          \includegraphics[width=\linewidth, trim=0 0 19cm 0, clip=true]{rotational_all.eps}
          \caption
          {
              Accumulated population of rotational states in a 40GHz wide window (blue) as a function of the rotational shift (running window).
              The X-axis is the rotational shift of Eq. \eqref{eq:rotational_broadening} of the corresponding rotational state $J$, and the Y-axis is the fractional population within the frequency window around this state.
              The red plot is a reproduction of the standard Boltzman distribution as given in figure \ref{fig:K2_rotational_accumulation}, 
              but now plotted against the rotational frequency shift. 
          }
        \label{fig:K2_rotational_broadening}
    \end{minipage}
\end{figure}

This inhomogeneous rotational broadening has immediate consequences on the threshold, indicating that a higher
density and pump energy (photon budget) would be required to cross threshold (of at least one coherent cluster).
Note however, that each coherent cluster is not necessarily composed of just a single $J$ state, and the
frequency bandwidth of a cluster can vary. Specifically, the width of a coherent cluster is set by the collisional
pressure bandwidth $\Delta_c\!=\!\gamma\!=\!1/T_2$, which may cover more than one $J$ state,
especially in the lower $J$ range, where the density of $J$ states is much higher
(the rotational energy is nearly quadratic in $J$). Since we can control the coherent collisional bandwidth
to some extent (by varying the temperature and by adding a buffer gas), there is a trade-off between the coherence
time (bandwidth of a single cluster) and the number of molecules within this cluster. For example,
assuming a coherence time of 25ps (40GHz), the highest populated clusters near the lower rotational
states $J\!<\!10$ hold $\sim\!6\%$ of the population, as shown in figure \ref{fig:K2_rotational_accumulation}.

Therefore, only 6\% of the molecular density is available for gain in a single coherent cluster
(depending on the available coherence time). A simple solution may be a 16-fold increase of the molecular density,
which does not compromise much of the available coherence time, since it requires to increase the vapor
temperature by only $\sim\!50^\circ \text{C}$ and the pressure to only 2-3Torr.
Yet, this will require also a 16-fold increase in pump energy, which may be problematic on the photon budget,
as all molecules in the beam will have to be pumped, regardless of whether they participate
in the oscillation or not. Thus, if pump energy is a limitation, a combination of measures can maintain
an affordable threshold condition. For example, to obtain threshold with an average pump power of 500mW,
at a molecular density of $10^{13}\text{cm}^{-3}$ (a 3-fold increase compared to table \ref{table:gain},
which sets pump absorption at $\sim\!15$\%), it is possible to: a) Reduce the cavity losses to a minimum of 1-2\%,
which will reduce the threshold by a factor of 2-3; b) Reuse the pump power in a low finesse of $F\!=\!3-4$,
which will enhance the pump power experienced by the molecules to 1.5-2W; and c) Use a lower repetition rate of 50MHz instead of 100MHz for both the pump
and the Raman cavity, which will increase the pump pulse energy to 30-40nJ at the same average power (another factor of 2).
In all, such a combination (or similar) allows to regain the missing gain due to rotational
broadening and maintain a reasonable threshold condition.

\subsection{\label{sec:expanded_simulation_results} Expanded Simulation Results}
Figure \ref{fig:simulation_results_near_threshold} shows simulation results for the temporal fields and populations with a molecular medium of $Li_2$ or $K_2$ dimers.
The two left columns show the amplification (small-signal) stage and the stable oscillation stage of $Li_2$ pumped slightly above threshold (a few percent).
Similarly, results for a slightly-above-threshold $K_2$ oscillator at stable oscillation are shown in the $3^{\text{rd}}$ column.
In the rightmost column, results at stable oscillation for $K_2$ oscillator pumped high above threshold are shown, where most of the population is dumped in a single
short pulse shortly after the pump pulse.

When pumped slightly above threshold, the accumulated field develops into a train of short pulses, matching the vibrational dynamics of the excited wave-packet,
which gradually dumps the entire excited population to a single vibrational final state.
However, if pumped high above threshold, a single dump pulse forms shortly after the pump pulse,
dumping most of the population very quickly, with no target selectivity.

In addition, two movie-clips are provided online to further illustrate the cavity buildup dynamics and threshold:
The first clip shows the target population evolution, broken into the different vibrational levels,
which demonstrates the exceptional selectivity of the transfer to a single vibrational target level
($>30dB$ compared to the next populated level), achieved very early during the cavity buildup,
well before substantial population is depleted from the excited wave-packet.
The second clip shows the intra cavity temporal intensity as it builds up in every cavity iteration,
demonstrating the threshold dynamics, which modifies the field once considerable population is depleted.
The clips show the intensity / population in log scale to enable observation of the dynamics over $>\!20$ orders of magnitude,
from the early noise-buildup to steady state operation.

\begin{figure*}[tbph]  
    \centering
      \includegraphics[width=\linewidth]{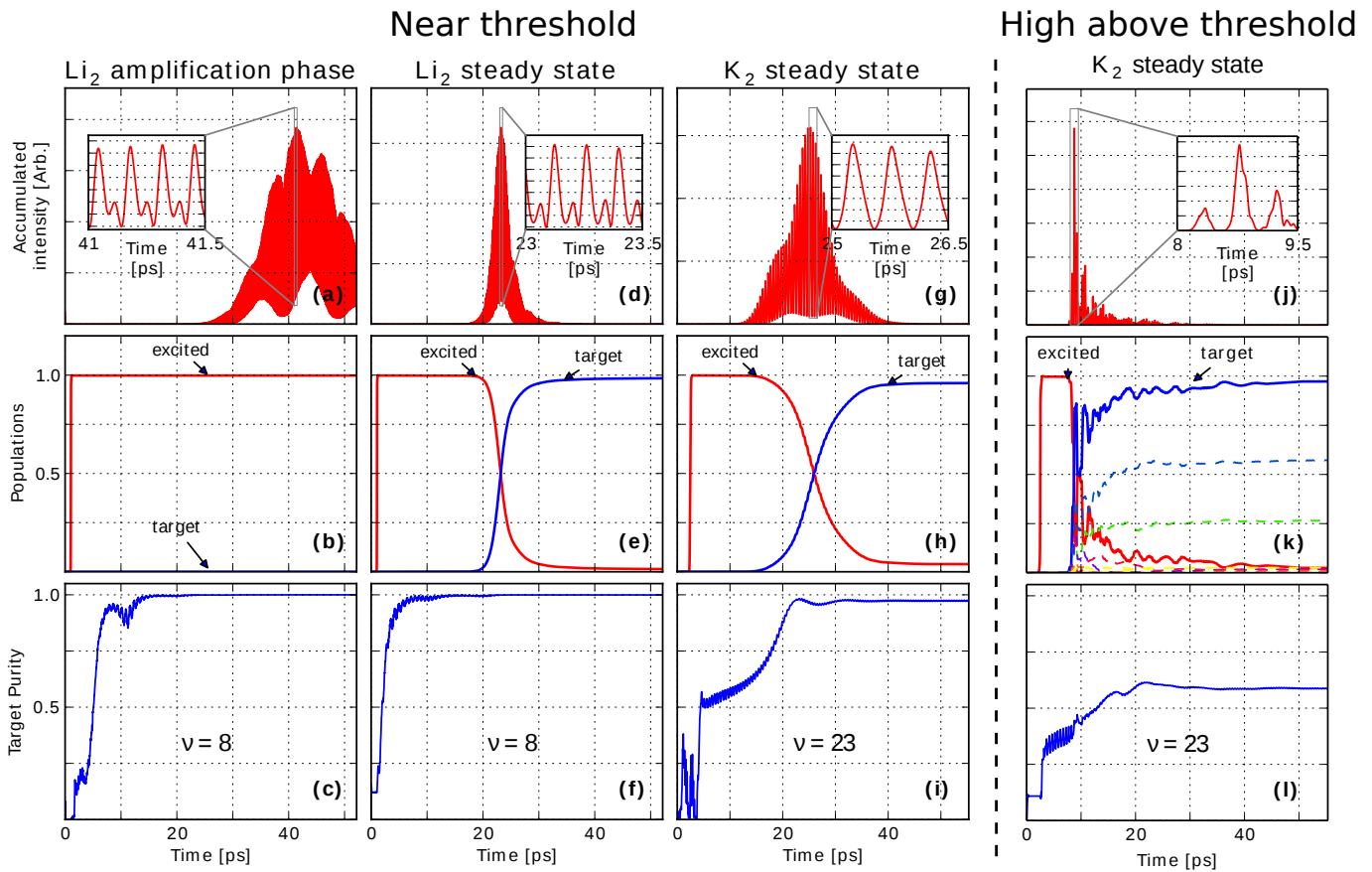}
      \caption
      {
          Simulation of the cavity dynamics (of a \emph{single coherent cluster}) for $Li_2$ and $K_2$ during 50ps
          following a particular pump pulse excitation. The results are arranged such that columns represent different stages in
          the cavity dynamics, and rows represent different types of data.
          The three left columns show results for an oscillator pumped \emph{slightly above} threshold (few percent),
          where the first column shows the intra-cavity field + populations at an early stage of the amplification,
          well before stable oscillation is reached ($Li_2$); the second column shows the result at a later stage,
          when stable oscillation is obtained ($Li_2$) and the third column shows results at stable oscillation
          for a medium of $K_2$. The rightmost column shows the stable oscillation for an oscillator pumped \emph{high above}
          threshold with a $K_2$ medium. The top row is the temporal intensity of the accumulated intra-cavity field,
          demonstrating rapid oscillation that is matched to the vibrational dynamics of the excited wave packet (see inset);
          Middle row - the temporal evolution of the total populations of the excited (red) and target (blue)
          wave-packets after the pump pulse, showing near complete dumping of the excited population,
          even for near-threshold pumping. Bottom - the overlap of the target state with the selected specific
          vibrational state $\nu=8$ for $Li_2$ and $\nu=23$ for $K_2$, showing practically $100\%$ selectivity of the
          transfer for $Li_2$ and $>98\%$ for $K_2$, when pumped near threshold.
          Note that high selectivity is achieved already during the amplification stage,
          well before appreciable population transfer is obtained. The $4^{\text{th}}$ column shows the same results for $K_2$ pumped high above threshold:
          the field (j) shows then a dominating single pulse which dumps most of the excited population
          to the target potential shortly after the pump excitation. The decomposition of the dumped population to 
          specific final vibrational states (dashed lines in (k)) shows that the target state is no longer a single vibration, 
          which is also reflected in the reduced selectivity of target state $\nu\!=\!23$ (l). Note that the field plots (a), (d), (g) and (j)
          are not on the same scale, as the accumulated fields in the stable oscillation stage are much higher.
          Furthermore, the population plots (solid lines) (b), (e), (h) and (k) are normalized to the maximal
          population of the excited wave packet $n(0)$ in each case, itself being a few percent of the original population of the input ground state.
      } \label{fig:simulation_results_near_threshold}
\end{figure*}

\subsubsection{\label{subsec:extraction_of_wave_packet_dynamics} Extraction of the excited wave-packet dynamics from the emitted field}
We can now show in detail how our major goal -
\emph{to read out the excited wave-packet dynamics from the emitted field}, can be realized. Since the target state is populated by a
single vibration $\nu_f$, the wave-packet dynamics is fully reflected in the time variation of
the molecular dipole $\braket{\nu_{f}|\psi_{e}(t)}$, and since the cavity field of figure \ref{fig:simulation_results_near_threshold}
is dominated by the coherent dipole term of Eq. \ref{eq:single_molecule_em_field}
$\bra{ \psi_{g}} U^2 \ket{\psi_{e}}$, the emitted field is practically a duplicate of the dipole.
Figure \ref{fig:li2_wave_packet_dynamics} shows the (normalized) Fourier transform of the molecular dipole moment (blue)
and the accumulated field (green) in the steady state of the oscillator, indicating that the
two are practically identical,
and perfectly reconstruct the eigen-frequencies of the excited vibrational components of the wave packet (red circles) \emph{and their phase},
with regard to the winning vibrational target mode $\nu_f$.
\begin{figure*}[tbph]  
    \centering
      \includegraphics[width=\linewidth]{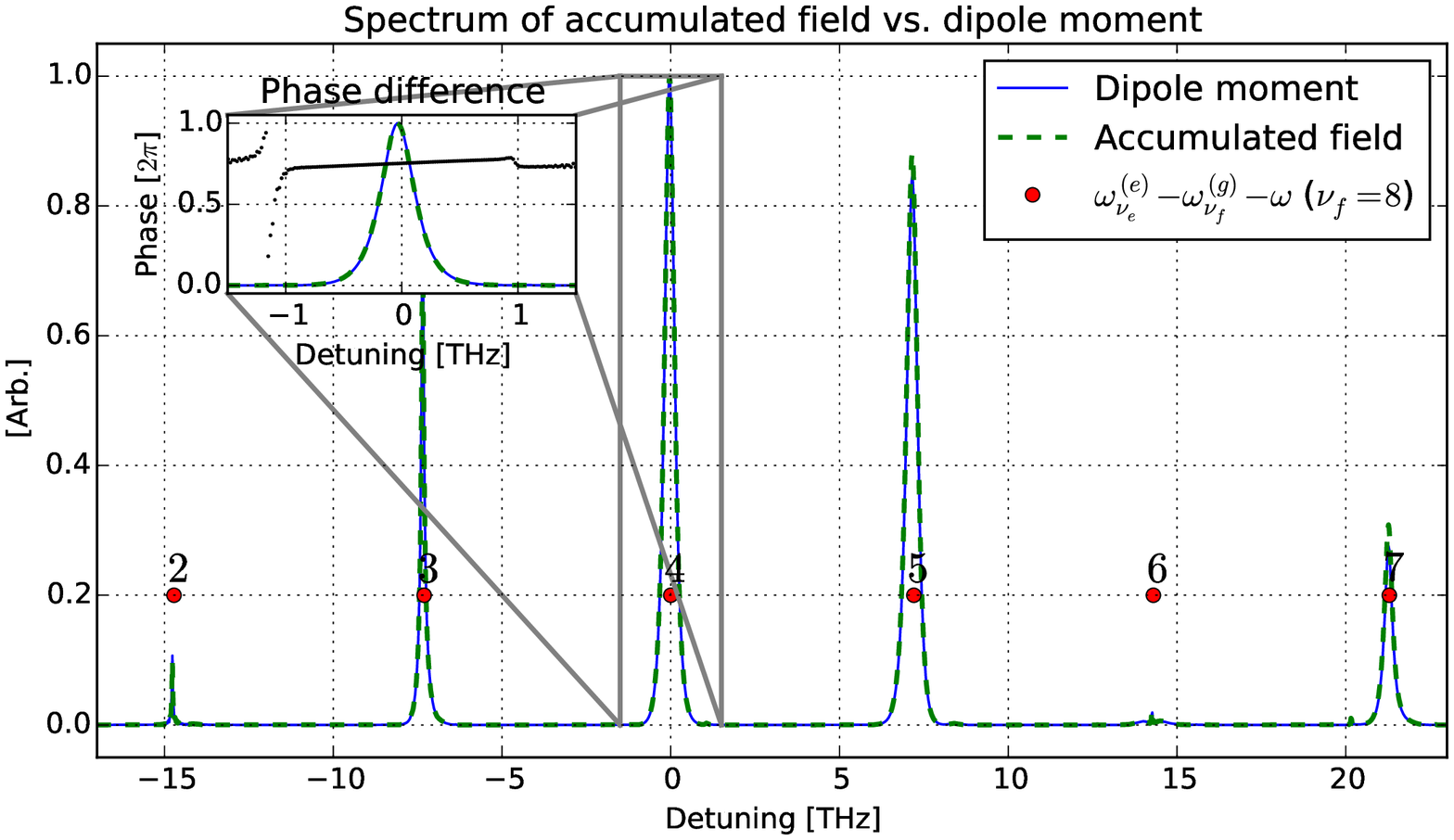}
      \caption
      {
               Extraction of wave-packet modes from the accumulated cavity field.
               The main graph shows the spectrum of the accumulated field (green) and the molecular dipole moment (blue)
               in the steady state of an oscillator based on a $Li_2$ dimer, in which an excited wave packet is centered
               around excited vibration $\nu_e=4$ and is dumped to a single target mode $\nu_f = 8$.
               The graphs of the cavity field and the molecular dipole are practically indistinguishable.
               The eigen-energies of the excited state $\omega^{(e)}_{\nu_e} - \omega^{(g)}_{\nu_f} - \omega$ are shown as red circles,
               where the center $\omega$ is given
               by $\omega = \omega^{(e)}_{4} - \omega^{(g)}_{8}$.
               The inset shows a zoom-in on the center vibrational line, including the relative phase between the molecular dipole and
               the emitted field (dotted black), where a nearly constant phase relation is observed.
               Thus, measuring the accumulated field and its phase provides a direct
               observation of the originating dipole and the dynamics of the excited wave packet.
      }
    \label{fig:li2_wave_packet_dynamics}
\end{figure*}

\subsubsection{\label{subsec:pump_shaping_and_target_mode_selection} Pump shaping and target mode selection}
As mentioned across the paper, the final state of the Raman transfer is not selected a-priori,
but rather dictated by the intra-cavity mode competition. However, this mode competition can be steered
towards a desired target state by shaping of the excited wave-packet.
Specifically, effective control is obtained by shaping the pump spectrum to excite within the wave-packet
only components with high Franck-Condon overlap to a \emph{designated} target state $\nu_f$ \cite{PhysRevLett.98.113004,PhysRevLett.101.023601}:
\begin{align}
    E(\omega^{(e)}_{\nu_e}  - \omega^{(g)}_{\nu_f}   )
    & \propto
        \text{FC}^{e}_{f}
    =
        \braket{ \phi^{(e)}_{\nu_e} | \phi^{(g)}_{\nu_f} }
\end{align}

This was demonstrated in cavity simulations with a gain medium of $Li_2$, where the produced cavity field was shown to
selectively dump the excited population to designated target states $\nu_f = 8, 9, 10$, just by shaping the pump spectrum with no other change in cavity parameters
(figure \ref{fig:pump_shaping} shows the relevant pump spectra), achieving the same level of selectivity as in figure
\ref{fig:simulation_results_near_threshold} (plots (c) and (f)).

\begin{wrapfigure}{r}{0.5\textwidth}
    \centering
      \includegraphics[width=\linewidth]{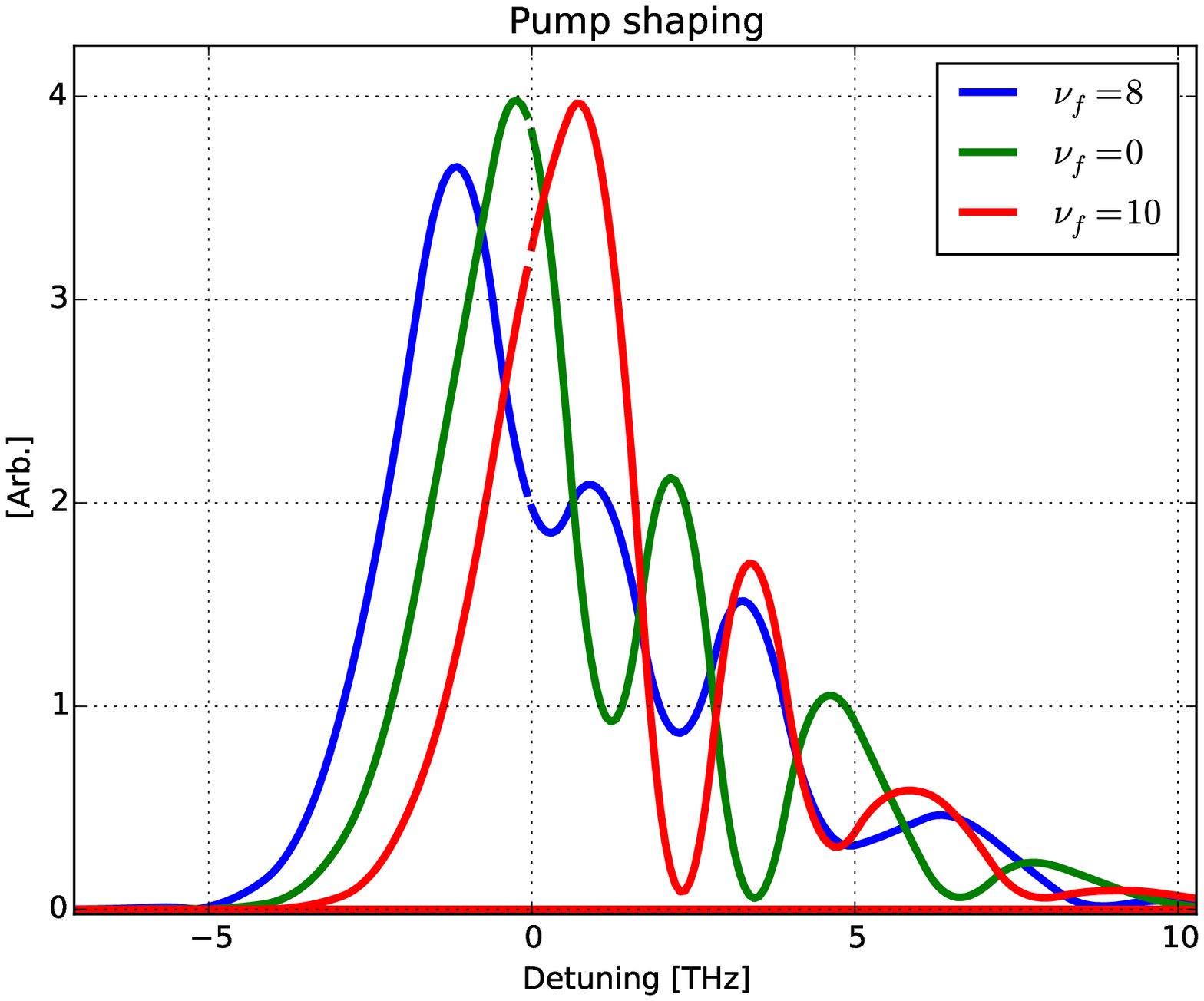}
      \caption
      {
          Steering the target state by shaping the pump spectrum in $Li_2$ for dumping into target modes $\nu_f=8, 9$ and $10$.
          The spectral peaks correspond to the states of maximal Franck-Condon overlap between the selected target mode and the range of excited modes around $\nu_e\!=\!4$.
      }
    \label{fig:pump_shaping}
\end{wrapfigure}
\clearpage
\end{widetext}

\clearpage
\bibliography{crl}

\end{document}